# COLOR-SINGLET CLUSTERS IN SYSTEMS COMPOSED OF SIX QUARKS

S. Mickevičius, G. P. Kamuntavičius

*Vytautas Magnus University, Donelaičio 58, 3000 Kaunas, Lithuania*

Saulius_Mickevicius@fc.vdu.lt

We present an investigation of six quarks system kinematics, independent on quark-quark interactions. There has been created and investigated basis of six quarks antisymmetrical and translationally invariant functions, applying formalism with spin, isospin, and color degrees of freedom and many-particle harmonic oscillator (HO) functions, dependent on Jacobian variables. We have also investigated expansions of realistic wave-functions in this basis. There are given analytical expressions for matrix elements of three and six particle antisymmetrizers and probability to find two colorless hadrons in six quarks system.

## 1. Introduction

The understanding of the interaction between two nucleons remains a fundamental problem in nuclear physics. There has been done a huge work for exploring hadrons structure and properties [1], but it is still unclear how the nuclear forces arise from the fundamental quark-quark interactions [2]. Starting from the common belief that quantum chromodynamics is the underlying theory of the strong interaction, one is entitled to demand that hadron physics has been formulated in terms of the basic quark degrees of freedom. Only in this way we will be able to describe the hadron structure and hadron-hadron interaction in consistent framework. Let us choose constituent quarks as the basic degrees of freedom.

As usual, solution of this problem is based on resonating group method. The first problem in such a case is proper symmetry and translational invariance of corresponding wave-functions. However, the application of usual simplifications for the radial functions of quarks and approximations for kernel are far from being complete and optimal. The more or less correct consideration of this problem [3] was performed only for configurations, corresponding to the lowest excitations. Only after modification of the approach [4,5], there is a possibility to solve this problem.

## 2. Antisymmetrization of quark clusters

We can look at a deuteron as a colorless (U=0) six quarks combination with total spin S=1 and isospin T=0. A complete antisymmetrizer of the six fermions has 6! elements. It can be written in such a form:

$$A_{1,2,3,4,5,6} = X_{1,2,3;4,5,6} A_{1,2,3} A_{4,5,6}, \qquad (1)$$

where $A_{1,2,3}$ and $A_{4,5,6}$ are antisymmetrizers of first and second three-quark clusters, and $X_{1,2,3;4,5,6}$ is antisymmetrizer of quarks from different clusters.

First of all, we antisymmetrized exchanges of quarks 1 and 2, as the restriction of two-quark channel quantum numbers by the condition

$$(-1)^{\bar{l}+\bar{s}+\bar{u}+\bar{t}} = -1, \qquad (2)$$

where $\bar{l}, \bar{s}, \bar{u}, \bar{t}$ are angular moment, spin, color and isospin of two-quarks. Three quarks antisymmetrizer, when 1 and 2 quarks are antisymmetrized and 3 is not antisymmetrized to the exchanges 1↔3 and 2↔3, equals:

$$X_{1,2;3} = \frac{1}{3}(1 - P_{13} - P_{23}). \qquad (3)$$

Due to the antisymmetry with respect to the exchanges 1↔2, the matrix elements of the antisymmetrizer $X_{1,2;3}$ can be evaluated simply as

$$\langle X_{1,2;3} \rangle = \frac{1}{3}(1 - 2\langle P_{23} \rangle). \qquad (4)$$

Its matrix element can be expressed in a straightforward way:

$$\langle P_{23} \rangle = [\bar{s}, \bar{s}', \bar{u}, \bar{u}', \bar{t}, \bar{t}']^{1/2} \begin{Bmatrix} 1/2 & 1/2 & \bar{t}' \\ 1/2 & t & \bar{t} \end{Bmatrix} \begin{Bmatrix} 1/2 & 1/2 & \bar{s}' \\ 1/2 & s & \bar{s} \end{Bmatrix} \begin{Bmatrix} 1 & 1 & \bar{u}' \\ 1 & u & \bar{u} \end{Bmatrix} \langle \varepsilon\lambda, \bar{e}\bar{l} : l \mid \varepsilon'\lambda', \bar{e}'\bar{l}' : l \rangle_{\frac{1}{3}},$$

where

$$[\bar{s}]^{1/2} = \sqrt{2\bar{s}+1},$$

and

$$\langle \varepsilon\lambda, \bar{e}\bar{l} : l \mid \varepsilon'\lambda', \bar{e}'\bar{l}' : l \rangle_{\frac{1}{3}}$$

is the general HO bracket for two particles [5] with mass ratio 1/3. The expression can be derived by examining the action of $P_{23}$. That operator changes the state

$$\left| \varepsilon\lambda(\vec{\xi}_1), \bar{e}\bar{l}(\vec{\xi}_2), l \right\rangle$$

to

$$\left|\varepsilon\lambda(\vec{\xi}_1'),\bar{e}\bar{l}(\vec{\xi}_2'),l\right\rangle.$$

The primed Jacobi coordinates can be expressed as an orthogonal transformation of the unprimed ones. Consequently, the HO wave functions, depending on the primed Jacobi coordinates can be expressed as an orthogonal transformation of the original HO wave functions. Elements of the transformation are the Talmi-Moshinsky HO brackets for two particles with the mass ratio $d$, with $d$ determined from the orthogonal transformation of the coordinates.

The antisymmetrizer $X$ is a projector satisfying:

$$X X = X.$$

When it is diagonalized in the three quarks basis, its eigenvectors span two eigenspaces. One, corresponding to the eigenvalue **1**, is formed by physical, completely antisymmetrized states and other, corresponding to the eigenvalue **0**, is formed by spurious states. There are about twice as many spurious states as the physical ones.

The resulting antisymmetrized states can be classified and expanded in terms of the original basis as follows:

$$\Psi_\gamma^{elsut}(\vec{\xi}_1,\vec{\xi}_2,\sigma_1,\sigma_2,\sigma_3,\rho_1,\rho_2,\rho_3,\tau_1,\tau_2,\tau_3) = $$
$$= \sum_{\varepsilon\lambda e\bar{e}\bar{l}s\bar{u}\bar{t}} \phi_{\varepsilon\lambda e\bar{e}\bar{l}s\bar{u}\bar{t}}^{elsut}(\vec{\xi}_1,\sigma_1,\rho,\tau_1;\vec{\xi},\sigma_2,\sigma_3,\rho_2,\rho_3,\tau_2,\tau_3) \times F_{\varepsilon\lambda e\bar{e}\bar{l}s\bar{u}\bar{t},\gamma}^{elsut}. \quad (5)$$

We introduced an additional quantum number $\gamma$ that distinguishes states with the same set of three-quarks quantum numbers *elsut*, $\gamma=1, 2, \ldots, r$ with $r$ the total number of antisymmetrized states for a given HO quantum numbers *el*, spin *s*, color *u*, and isospin *t*. It can be obtained from computing the trace of the antisymmetrizer $X$. $F$ are eigenvectors, corresponding to the eigenvalue **1**.

Antisymmetrizer for six quarks

$$X_{1,2,3;4,5,6} = \frac{1}{20}\left(1 - \sum_{i=1}^{3}\sum_{j=4}^{6} P_{ij} + \sum_{i_1<i_2=1}^{3}\sum_{j_1<j_2=4}^{6} P_{i_1 j_1} P_{i_2 j} - P_{14} P_{25} P_{36}\right) = $$
$$= \frac{1}{10}\left(1 - \sum_{i=1}^{3}\sum_{j=4}^{6} P_{ij}\right)\frac{1}{2}(1 - P_{14} P_{25} P_{36}) = X_{(1,2,3;4,5,6)} A_{(1,2,3;4,5,6)} \quad (6)$$

split down into two parts**.** The last operator antisymmetrize clusters. If all quantum numbers of the first and the second clusters are equal, two-clusters channel quantum numbers are restricted by the condition

$$(-1)^{l+\bar{L}+S+U+T} = -1 \quad (7)$$

where $S, U, T, \bar{L}$ are total spin, color, isospin and angular moment of the two clusters and $l$ is angular moment of clusters relative motion. If quantum numbers of clusters are not equal, the eigenvectors of the antisymmetrizer are:

$$\begin{pmatrix} \dfrac{1}{\sqrt{2}} \\ -\dfrac{1}{\sqrt{2}}(-1)^{l+l_1+l_2+\bar{L}+s_1+s_2+S+t_1+t_2+T} \end{pmatrix}. \qquad (8)$$

The last part of antisymmetrizer $X_{(1,2,3;4,5,6)}$ from (6), which exchanges quarks between the clusters, due to the antisymmetry with respect to the exchanges inside the clusters, can be evaluated simply as

$$\langle X_{(1,2,3;4,5,6)} \rangle = \frac{1}{10}(1 - 9\langle P_{14} \rangle). \qquad (9)$$

Its matrix element can be expressed in a straightforward way:

$$\langle P_{14} \rangle = \langle P_{14}^{spin} \rangle \langle P_{14}^{color} \rangle \langle P_{14}^{isospin} \rangle \langle P_{14}^{orb} \rangle, \text{ where}$$

$$\langle P_{14}^{spin} \rangle = \partial_{\bar{s}_1,\bar{s}_1'} \partial_{\bar{s}_2,\bar{s}_2'} (-1)^{\bar{s}_1+\bar{s}_2-s_2+s_2'} [s_1,s_2,s_1',s_2']^{1/2} \begin{Bmatrix} 1/2 & \bar{s}_1 & s_1 \\ \bar{s}_2 & 1/2 & s_2 \\ s_2' & s_1' & 1 \end{Bmatrix}, \qquad (10)$$

$$\langle P_{14}^{color} \rangle = \partial_{\bar{u}_1,\bar{u}_1'} \partial_{\bar{u}_2,\bar{u}_2'} \partial_{u_1,u_2} \partial_{u_1',u_2'} [u_1,u_1']^{1/2} \begin{Bmatrix} 1 & \bar{u}_1 & u_1 \\ 1 & \bar{u}_2 & u_1' \end{Bmatrix}, \qquad (11)$$

$$\langle P_{14}^{isospin} \rangle = \partial_{\bar{t}_1,\bar{t}_1'} \partial_{\bar{t}_2,\bar{t}_2'} \partial_{t_1,t_2} \partial_{t_1',t_2'} [t_1,t_1']^{1/2} \begin{Bmatrix} 1/2 & \bar{t}_1 & t_1 \\ 1/2 & \bar{t}_2 & t_1' \end{Bmatrix}, \qquad (12)$$

$$\langle P_{14}^{orb} \rangle = (-1)^l [l_1,l_2,l_1',l_2',\bar{L},\bar{L}']^{1/2} \partial_{(\bar{e}\bar{l})_1,(\bar{e}\bar{l})_1'} \partial_{(\bar{e}\bar{l})_2,(\bar{e}\bar{l})_2'} \sum_{\Lambda,\Lambda',\bar{\bar{L}},\bar{\Lambda}} [\Lambda,\Lambda',\bar{\bar{L}},\bar{\Lambda}] \times$$

$$\times \begin{Bmatrix} l & \Lambda & \bar{\Lambda} \\ \bar{\bar{L}} & L & \bar{L} \end{Bmatrix} \begin{Bmatrix} l' & \Lambda' & \bar{\Lambda} \\ \bar{\bar{L}} & L & \bar{L}' \end{Bmatrix} \begin{Bmatrix} \lambda_1 & \bar{l}_1 & l_1 \\ \lambda_2 & \bar{l}_2 & l_2 \\ \Lambda & \bar{\bar{L}} & \bar{L} \end{Bmatrix} \begin{Bmatrix} \lambda_1' & \bar{l}_1 & l_1' \\ \lambda_2' & \bar{l}_2 & l_2' \\ \Lambda' & \bar{\bar{L}} & \bar{L}' \end{Bmatrix} \times$$

$$(13)$$

$$\times \sum_{\substack{E_1\Lambda_1 E_2\Lambda_2 \\ \bar{\Lambda}\mu\nu}} (-1)^\mu [\tilde{\Lambda}] \begin{Bmatrix} l & \nu & \tilde{\Lambda} \\ \Lambda_1 & \bar{\Lambda} & \Lambda \end{Bmatrix} \begin{Bmatrix} l' & \Lambda_2 & \tilde{\Lambda} \\ \Lambda_1 & \bar{\Lambda} & \Lambda' \end{Bmatrix} \langle E_1\Lambda_1, E_2\Lambda_2 : \Lambda' | \varepsilon_1'\lambda_1', \varepsilon_2'\lambda_2' : \Lambda' \rangle_1 \times$$

$$\times \langle E_1\Lambda_1, \mu\nu : \Lambda | \varepsilon_1\lambda_1, \varepsilon_2\lambda_2 : \Lambda \rangle_1 \langle E_2\Lambda_2, e'l' : \tilde{\Lambda} | \mu\nu, el : \tilde{\Lambda} \rangle_{1/8}.$$

Spectrum of antisymmetrizers $X_{(1,2,3;4,5,6)}$ matrix gives us completely antisymmetrical and translationally invariant six-quark system wave-functions in HO basis.

### 3. Conclusions

Using above results, we have developed a Fortran code for the calculation of three-quark antisymmetrizer (3) and six-quark antisymmetrizer (7) matrixes. Calculation have been done for HO energies $E$ up to 6 quanta. The results are presented in Table 1. The main result of the calculation is such, that probability to find two colorless hadrons in six quark system is stable and independent on HO energy, and equals 1/5.

Table 1.

| HO quantum numbers | Matrix dimension | Number of colorless states | Probability of two colorless hadrons in six quarks system | Calculations time |
|---|---|---|---|---|
| E=0, L=0 | 14 | 2 | 1/5 | 1 s |
| E=2, L=0 E=2, L=2 | 152 | 14 | 1/5 | 90 s |
| E=4, L=0 | 1168 | 94 | 1/5 | 10 min |
| E=4, L=2 | 2123 | 161 | 1/5 | 30 min |
| E=6, L=0 | 6448 | 478 | 1/5 | 20 h |
| E=6, L=2 | 16414 | 1645 | 1/5 | 120 h |